\documentclass[twocolumn,floatfix,aps,superscriptaddress]{revtex4}
\usepackage{graphicx}
\usepackage{amsmath}
\usepackage{amssymb}
\usepackage{bm}

\usepackage{color}

\begin{document}

\title{Adaptive weight estimator for quantum error correction\\
in a time-dependent environment}
\author{S. T. Spitz}
\affiliation{Instituut-Lorentz, Universiteit Leiden, P.O. Box 9506, 2300 RA Leiden, The Netherlands}
\author{B. Tarasinski}
\affiliation{QuTech, Delft University of Technology, P.O. Box 5046, 2600 GA Delft, The Netherlands}
\author{C. W. J. Beenakker}
\affiliation{Instituut-Lorentz, Universiteit Leiden, P.O. Box 9506, 2300 RA Leiden, The Netherlands}
\author{T. E. O'Brien}
\affiliation{Instituut-Lorentz, Universiteit Leiden, P.O. Box 9506, 2300 RA Leiden, The Netherlands}

\date{December 2017}
\begin{abstract}
Quantum error correction of a surface code or repetition code requires the pairwise matching of  error events in a space-time graph of qubit measurements, such that the total weight of the matching is minimized. The input weights follow from a physical model of the error processes that affect the qubits. This approach becomes problematic if the system has sources of error that change over time. Here we show how the weights can be determined from the measured data in the absence of an error model. The resulting adaptive decoder performs well in a time-dependent environment, provided that the characteristic time scale $\tau_{\rm env}$ of the variations is greater than $\delta t/\bar{p}$, with $\delta t$ the duration of one error-correction cycle and $\bar{p}$ the typical error probability per qubit in one cycle.
\end{abstract} 
\maketitle

\section{Introduction}
\label{intro}

To execute algorithms on a quantum computer, one must prevent the accumulation of errors by monitoring and correcting them in control hardware. The monitoring is made possible by a nonlocal encoding of the quantum information in a redundant set of qubits, allowing for repeated measurements via auxiliary (ancilla) qubits without collapsing and destroying the quantum superposition of the logical degrees of freedom~\cite{Lid13,Ter15}. Parity-check measurements produce strings of bits, the so-called error syndrome, that must be decoded to infer the correction which should be applied to the logical qubits.

For an important class of error correcting codes, the syndrome identifies the end points of an error chain in a space-time graph of ancilla measurements. (See Fig.~\ref{repcode}.) The dimensionality of space can differ; it equals 1 in the repetition code~\cite{Sho95,Ste96}, 2 in the surface code~\cite{Bra98,Fow12a,Kel15}, and 3 for topological cluster states~\cite{Rau07}. The identification is not unique: there is in general no unique way to construct a chain of error events consistent with a given syndrome (the decoding problem). One approach to decoding refers to the optimization problem of minimum-weight perfect matching on a graph, which may be solved by the ``blossom'' algorithm~\cite{Edm65,Fow15} in polynomial time. The blossom decoder is sub-optimal~\cite{Fow13,Del14,Hei16,Bai17}, but it performs sufficiently well for current quantum hardware to achieve the fault-tolerance threshold~\cite{Obr17}.

The weights that govern the optimization problem can be readily obtained if one has a calibrated model of the sources of error in the system~\cite{Fow12}. Such an error model may not be available, and moreover the error rates may vary in time during the quantum computation. This complication has motivated the search for an adaptive decoder, that would infer the weights from the syndrome without requiring updates of the error model~\cite{Fow14,Com14,Fuj14,Ors16,Huo17}. Since the syndrome depends nonlinearly on the weights, this inversion problem is nontrivial --- a  recent approach~\cite{Huo17} employs a machine learning algorithm to learn the weights from the measured data.

Here we show that the inversion can be actually carried out by purely algebraic means: The covariance of measurements on pairs of ancillas exactly determines the weight of their matching. We demonstrate the method on the repetition code with time-dependent error rates.

\section{Quantum error correction and the repetition code}
\label{QECrepetition}

To set the stage, we summarize the elements of quantum error correction~\cite{Dev13,Ter15} that we need in what follows. The expert reader may skip this section.

A quantum error correcting code stores quantum information nonlocally in an array of physical qubits, such that it is protected from local errors (bit flips or sign flips). The encoded state $|\psi\rangle$ evolves for a cycle time $\delta t$, after which a set of `stabilizer' measurements is carried out. The stabilizers project $|\psi\rangle$ onto a state $|\psi'\rangle$ that may differ from $|\psi\rangle$ if an error occured during the cycle. The outcome of the stabilizer measurements, called the syndrome, identifies the error and allows for a correction. It is crucial that the stabilizer measurements do not measure the degrees of freedom of $|\psi\rangle$ in which the relevant quantum information is stored, otherwise this information will be lost upon projection.

\begin{figure}[tb]
\centerline{\includegraphics[width=1\linewidth]{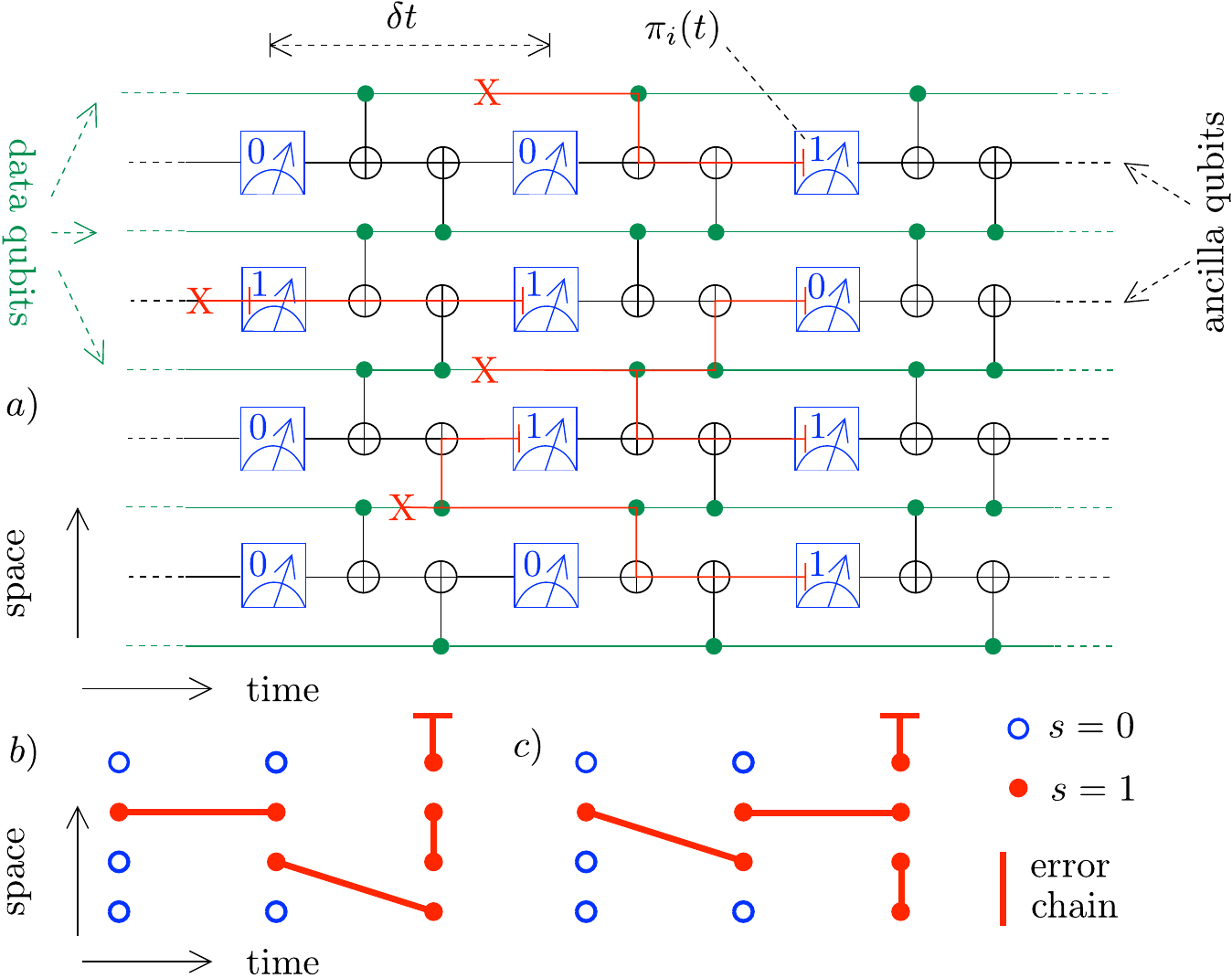}}
\caption{\textit{Panel a:} Space-time circuit of a $d=5$ quantum repetition code. Five data qubits (green) are entangled with four ancilla qubits (black) by means of a {\tt CNOT} gate ($\oplus$). The ancillas are measured at the end of each cycle (blue boxes, spaced by $\delta t$, with measurement outcomes $\pi_i(t)$, $i=1,2,3,4$). A bit flip (red X) produces an error chain (red line) with end points on an ancilla measurement or on the boundary of the array. \textit{Panel b:} Syndrome ${s}_i(t)=\pi_i(t)\oplus\pi_i(t-2\delta t)$ corresponding to the error events in panel a). The measurements that are connected by an error chain can be separated in space, in time, or in both space and time. \textit{Panel c:} Alternative matching consistent with the same syndrome. The minimum weight decoder associates a weight to each error chain and finds the matching with the smallest total weight.
}
\label{repcode}
\end{figure}

The simplest example of error correction via stabilizer measurement is a one-dimensional array, in which a logical qubit is encoded into $d$ data qubits via $|0\rangle_{\rm L} = |00\cdots0\rangle$, $|1\rangle_{\rm L}=|11\cdots1\rangle$. In a classical setting, this would correspond to a distance-$d$ repetition code, for which one would compare the value of adjacent bits to identify up to $(d-1)/2$ bit flips. A quantum parity check achieves this goal without collapsing the superposition $|\psi\rangle=a_0|0\rangle_{\rm L}+a_1|1\rangle_{\rm L}$ onto the state $|0\rangle_{\rm L}$ or $|1\rangle_{\rm L}$. The parity-check measurements are performed on $d-1$ ancilla qubits, which are entangled with pairs of data qubits (see Fig.\ \ref{repcode}). Each ancilla measures the stabilizer operator $Z_iZ_{i+1}$ ($i=1,2,\ldots d-1$, with $Z\equiv\sigma_z$ a Pauli matrix). The stabilizer does not distinguish between the states $|0\rangle_{\rm L}$ and $|1\rangle_{\rm L}\rangle$ and thus preserves their quantum superposition. A bit flip error of qubit $j$ ($X_j$ error) is detected by the stabilizer measurements $Z_j Z_{j+1}$ and $Z_{j-1}Z_{j}$, which change their value from $+1$ to $-1$. A decoder may infer the underlying error from this signature and correct it without needing to measure qubit $j$ itself (which would collapse the state).

The stabilizer measurements in cycle $t$ form a binary parity-check vector $\vec{\pi}(t)$. The error syndrome
\begin{equation}
\vec{s}(t)=\vec{\pi}(t)+\vec{\pi}(t-2\delta t)\label{spidef}
\end{equation}
is defined such that an error event is signaled by a nonzero element $s_i(t)=1$. A single error event is not sufficient to diagnose an error, as $Z_i Z_{i+1}$ would trigger an error event for either $X_i$ or $X_{i+1}$. To identify which qubit flipped we match pairs of error events. As indicated in Fig.~\ref{repcode}, the match can be between error events that are separated in both space and time, at the end points of an error chain from ancilla $i_0$ at time $t$ to ancilla $j_0$ at time $t+n\delta t$. The error chain may terminate at the boundary of the lattice (corresponding to errors on the boundary data qubits), so some error events may remain unmatched.

This simple description to detect bit flips in a repetition code can be extended to the detection of both bit flips and phase flips ($X_i$ and $Z_i$ errors) and by encoding in 2D and 3D (surface codes and topological cluster states). The generic feature of this class of stabilizer codes is that the decoding entails the pairwise matching of error events in a space-time graph. The method of adaptive quantum error correction presented in the next section applies to this general setting, while for a demonstration we will return to the repetition code. 

\section{Weight inference from error syndromes}
\label{weightinference}

\subsection{Formulation of the inversion problem}
\label{formulation}
 
\begin{figure}[tb]
\centerline{\includegraphics[width=0.4\linewidth]{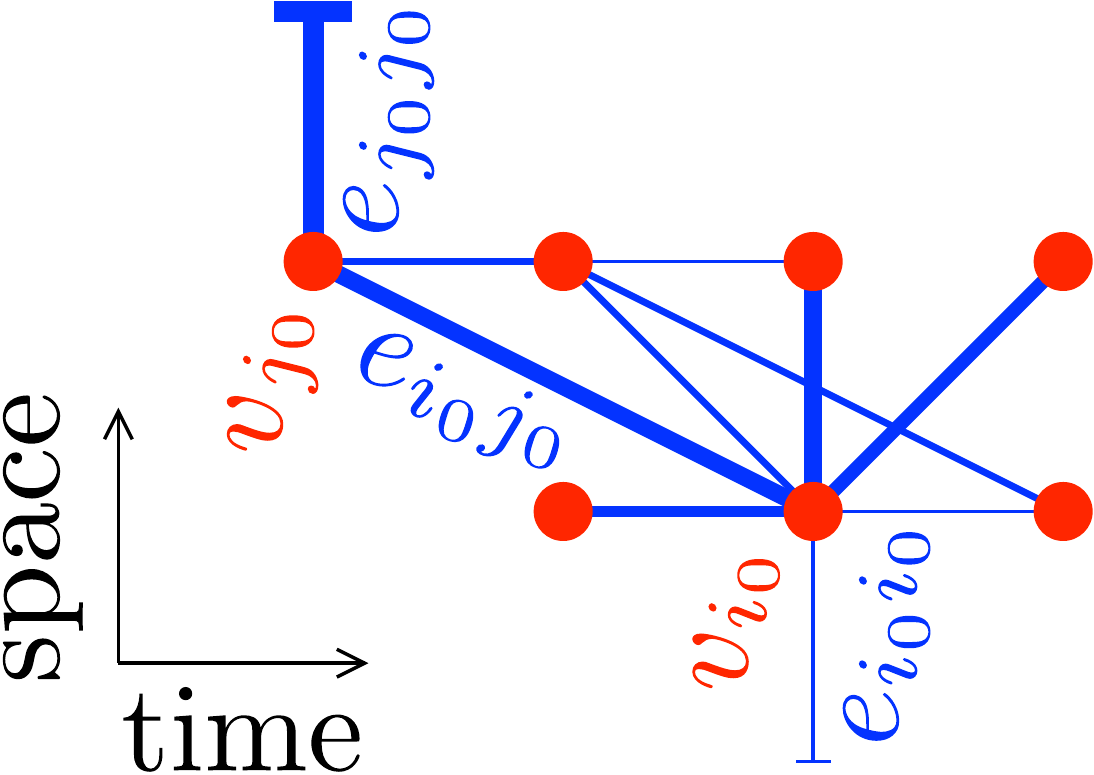}}
\caption{Space-time graph showing a pair of vertices $v_{i_0}$, $v_{j_0}$ connected by an edge $e_{i_0j_0}$. A few other vertices and connecting edges are also shown, as well as edges that connect a vertex to a boundary ($e_{i_0i_0}$ and $e_{j_0j_0}$). The line thickness of an edge $e_{ij}$ is proportional to the probability $p_{ij}$ that a single error affects the ancilla qubit measurements on vertices $i$ and $j$. We seek to determine these probabilities from measurements of the error syndrome. 
}
\label{fig_graph}
\end{figure}

We collect the binary output of the stabilizer measurements in the error syndrome $\vec{s}(t)$. The discrete time variable $t$ counts the error correction cycle and the elements of the vector $\vec{s}$ identify the ancilla qubits. For $N$ ancillas and $T$ cycles there are a total of $NT$ variables $v_i\in\{0,1\}$, arranged as vertices in a space-time graph. (See Fig.\ \ref{fig_graph}.) An error event corresponds to $v_i=1$, while $v_i=0$ if the ancilla has not detected an error.

The vertices are pairwise connected by undirected edges $e_{ij}\equiv e_{ji}\in\{0,1\}$ such that $e_{ij}=1$ with probability $p_{ij}$. We allow for $i=j$, the edge $e_{ii}$ connects a vertex to the boundary of the graph. We say that the edge is \textit{on} or \textit{off} depending on whether $e_{ij}=1$ or $0$. The state of a vertex depends on the edges according to
\begin{equation}
v_{i}=\tfrac{1}{2}\bigl[1-(-1)^{\sum_{j}e_{ij}}\bigr].\label{vieijrelation}
\end{equation}
Each edge that is \textit{on} toggles its vertex between the states $0$ and $1$, so that $v_i=1$ if an odd number of connecting edges is \textit{on}.

The $p_{ij}$'s are probabilities of a single-qubit error that correlates ancilla measurements $i$ and $j$. Correlations of ancilla measurements due to uncorrelated multiple-qubit errors are described by weights $w_{ij}$. The weight $w_{ij}$ for $i\neq j$ is determined from the $p$'s by following all paths through the graph from $i$ to $j$:
\begin{equation}
w_{ij}=-\ln\left(p_{ij}+\sum_n\sideset{}{'}\sum_{k_1,k_2,\ldots k_n}p_{ik_1}p_{k_1k_2}\cdots p_{k_nj}\right).
\end{equation}
The prime in the sum indicates that the path should not pass through the boundary ($i\neq k_1\neq k_2\cdots\neq k_n\neq j$). For a boundary weight $w_{ii}$ the path terminates on the boundary,
\begin{equation}
w_{ii}=-\ln\left(p_{ii}+\sum_n\sideset{}{'}\sum_{k_1,k_2,\ldots k_n}p_{ik_1}p_{k_1k_2}\cdots p_{k_nk_n}\right).
\end{equation}
These sums over error chains can be carried out in terms of the matrix $A_{ij}=(1-\delta_{ij})p_{ij}$ by matrix inversion \cite{Obr17},
\begin{equation}
e^{-w_{ij}}=\begin{cases}
[(1-A)^{-1}]_{ij}&\text{if}\;\;i\neq j,\\
\sum_{k}[(1-A)^{-1}]_{ik}p_{kk}&\text{if}\;\;i= j.
\end{cases}\label{weightprelation}
\end{equation}

Given a set of error events ${\cal V}=\{v_i|v_i=1\}$ the minimum-weight perfect matching decoder searches for a subset ${\cal M}=\{e_{ij},w_{ij}\}$ of weighted edges such that each vertex in ${\cal V}$ is connected either to one other vertex in ${\cal V}$ or to the boundary, at minimal total weight $\sum_{\cal M}w_{ij}$. 

Modern implementations \cite{Fow15} of the blossom algorithm \cite{Edm65} solve this optimization problem efficiently given an error model: A physical model for qubit errors from which the error probabilities $p$ and hence the weights $w$ can be calculated. Here we consider the opposite approach: can we infer the weights from the measured error syndromes, without an underlying error model? This is an inversion problem for Eq.\ \eqref{vieijrelation}, where we seek to reconstruct the statistics of the edges $e_{ij}$ from the measured statistics of the vertices $v_{i}$. The inversion is possible, in spite of the nonlinearity of Eq.\ \eqref{vieijrelation}, as we now show.

\subsection{Solution for edges connecting pairs of vertices}
\label{solutionlinked}

We first consider a pair of distinct vertices $i_0\neq j_0$, connected by an edge $e_{i_0j_0}$. (The case of a single vertex connected to the boundary will be dealt with later.) We define
\begin{equation}
E_{i_0\backslash j_0}=\tfrac{1}{2}\bigl[1+(-1)^{\sum_{j\neq j_0}e_{i_0 j}}\bigr].\label{Pi0j0}
\end{equation}
In words, $E_{i_0\backslash j_0}$ equals 1 or 0 depending on whether the vertex $i_0$ has an even or an odd number of connecting edges that are \textit{on} --- excluding the connection to vertex $j_0$. Note that the sum over $j$ includes $j=i_0$, it only excludes $j=j_0$.

We then rewrite Eq.\ \eqref{vieijrelation} for vertex $i_0$ as
\begin{equation}
v_{i_0}=e_{i_0j_0}E_{i_0\backslash j_0}+(1-e_{i_0j_0})(1-E_{i_0\backslash j_0}).\label{vi0def}
\end{equation}
Similarly, for vertex $j_0$ one has
\begin{equation}
v_{j_0}=e_{j_0i_0}E_{j_0\backslash i_0}+(1-e_{j_0i_0})(1-E_{j_0\backslash i_0}).\label{vj0def}
\end{equation}
Since $e_{i_0j_0}=e_{j_0i_0}=e^2_{i_0j_0}$, the product ({\tt AND}) of $v_{i_0}$ and $v_{j_0}$ equals
\begin{align}
v_{i_0}v_{j_0}={}&(1-e_{i_0j_0})(1-E_{i_0\backslash j_0}-E_{j_0\backslash i_0})\nonumber\\
&+E_{i_0\backslash j_0}E_{j_0\backslash i_0},\label{vi0vj0}
\end{align}
while the binary sum ({\tt XOR}) equals
\begin{align}
v_{i_0}\oplus v_{j_0}&\equiv v_{i_0}+ v_{j_0}\;\text{mod}\;2\nonumber\\
&=E_{i_0\backslash j_0}+E_{j_0\backslash i_0}-2E_{i_0\backslash j_0}E_{j_0\backslash i_0}.\label{vi0plusvj0}
\end{align}

By construction, all three variables $e_{i_0j_0}$, $E_{i_0\backslash j_0}$, and $E_{j_0\backslash i_0}$ are statistically independent. We denote the average by $\langle\cdots\rangle$, with 
\begin{equation}
\langle e_{i_0j_0}\rangle=p_{i_0j_0}\label{ei0j0average}
\end{equation}
by definition. The averages of the $E$'s are unknown, but they can be eliminated by combining the four equations \eqref{vi0def}--\eqref{vi0plusvj0}. We thus arrive at
\begin{align}
&p_{i_0j_0}(1-p_{i_0j_0})=\frac{\langle v_{i_0}v_{j_0}\rangle-\langle v_{i_0}\rangle\langle v_{j_0}\rangle}{1-2\langle v_{i_0}\oplus v_{j_0}\rangle}.\label{linkedp}
\end{align}
The left-hand-side is symmetric under the exchange $p_{i_0j_0}\leftrightarrow 1-p_{i_0j_0}$. We may safely assume that the error probabilities are $<1/2$, resulting in the probability
\begin{equation}
p_{i_0j_0}=\frac{1}{2}-\sqrt{\frac{1}{4}-\frac{\langle v_{i_0}v_{j_0}\rangle-\langle v_{i_0}\rangle\langle v_{j_0}\rangle}{1-2\langle v_{i_0}\oplus v_{j_0}\rangle}}.\label{linked}
\end{equation}
This is an exact relation between the probability of an edge and correlators of the pair of connected vertices. These correlators are measurable from the error syndrome, without any prior knowledge of the error model.

\subsection{Solution for boundary edges}
\label{solutionboundary}

The probability $p_{i_0i_0}$ of an edge $e_{i_0i_0}$ connecting the vertex $v_{i_0}$ to the boundary cannot be determined by a correlator, since there is nothing to correlate with. We do have access to the average
\begin{align}
\langle v_{i_0}\rangle&=1-p_{i_0i_0}-(1-2p_{i_0i_0})\langle E_{i_0\backslash i_0}\rangle,\label{vi0average}\\
E_{i_0\backslash i_0}&=\tfrac{1}{2}\bigl[1+(-1)^{\sum_{j\neq i_0}e_{i_0 j}}\bigr]\nonumber\\
&=\tfrac{1}{2}+\tfrac{1}{2}\prod_{j\neq i_0}(1-2e_{i_0j}).\label{Ei0i0def}
\end{align}
Using again the independence of the variables, we find
\begin{equation}
p_{i_0i_0}=\frac{1}{2}+\frac{\langle v_{i_0}\rangle-1/2}{\prod_{j\neq i_0}(1-2p_{i_0j})}.\label{boundary}
\end{equation}
So once the probabilities $p_{i_0j}$ for non-boundary edges are determined from Eq.\ \eqref{linked}, we can use Eq.\ \eqref{boundary} to obtain the probability of a boundary edge.

\section{Implementation of the adaptive decoder}

\subsection{Convergence in the large-time limit}

We test the adaptive decoder on the repetition code of Fig.~\ref{repcode}, for a bit-flip error model: at the end of a cycle of duration $\delta t$ each qubit $i$ is flipped independently with probability $\gamma_i$. The time-dependent density matrix of the quantum circuit is calculated using the  {\tt quantumsim} simulator of Ref.\ \onlinecite{Obr17}. 

We implement the blossom decoder without any prior knowledge of the error probabilities, using Eqs.~\eqref{linked} and~\eqref{boundary} to determine them from the measured syndrome data. We assume local sources of error and set $p_{ij}\equiv 0$ for ancilla measurements $i$ and $j$ that are not connected by any local error. In a nonlocal situation, \textit{e.g.} because of non-negligible crosstalk, a proliferation of negligibly small error probabilities can be avoided by setting $p_{ij}\equiv 0$ when the deviation from zero is statistically insignificant.

The adaptive decoder needs sufficient syndrome data in the training stage to estimate the probabilities. Since $p_{ij}$ is the mean of a Bernoulli random variable with variance $\sigma_{ij}^2=p_{ij}(1-p_{ij})$, the statistical uncertainty $\delta p_{ij}$ in the estimation after $N=t/\delta t$ error cycles is of order
\begin{equation}
\delta p_{ij}=N^{-1/2}\sqrt{p_{ij}(1-p_{ij})}.\label{eq:stderr}
\end{equation}
The requirement that $\delta p_{ij}\ll p_{ij}\ll 1$ implies that a minimum of 
\begin{equation}
N_{\rm min}\simeq 1/\bar{p}
\end{equation}
measurements are needed for a reliable estimation of error probabilities of typical magnitude $\bar{p}$. After the training stage the probabilities are inserted into Eq.\ \eqref{weightprelation} to determine the weights which are passed to the minimum-weight perfect matching (blossom) decoder for error correction.

As a figure of merit we introduce a testing stage after the training stage in which we calculate the probability $\epsilon_{\rm adaptive}(N)$ of a logical error per cycle using the adaptive decoder trained on $N$ rounds of data. The error rates are calculated following the method of Ref.~\onlinecite{Obr17}, measuring the average logical qubit fidelity over 100 cycles. The combination of training and testing is repeated a few hundred times to obtain an accurate value of $\epsilon_{\rm adaptive}(N)$. We compare this with the probability $\epsilon_{0}$ that would follow from a blossom decoder with pre-determined weights calculated from the error model. The relative error
\begin{equation}
\Delta=\epsilon_{\rm adaptive}/\epsilon_0-1\label{eq:reldecerr}
\end{equation}
measures how well the adaptive decoder has converged to the ideal blossom decoder.

\begin{figure}[tb]
\centerline{\includegraphics[width=0.9\linewidth]{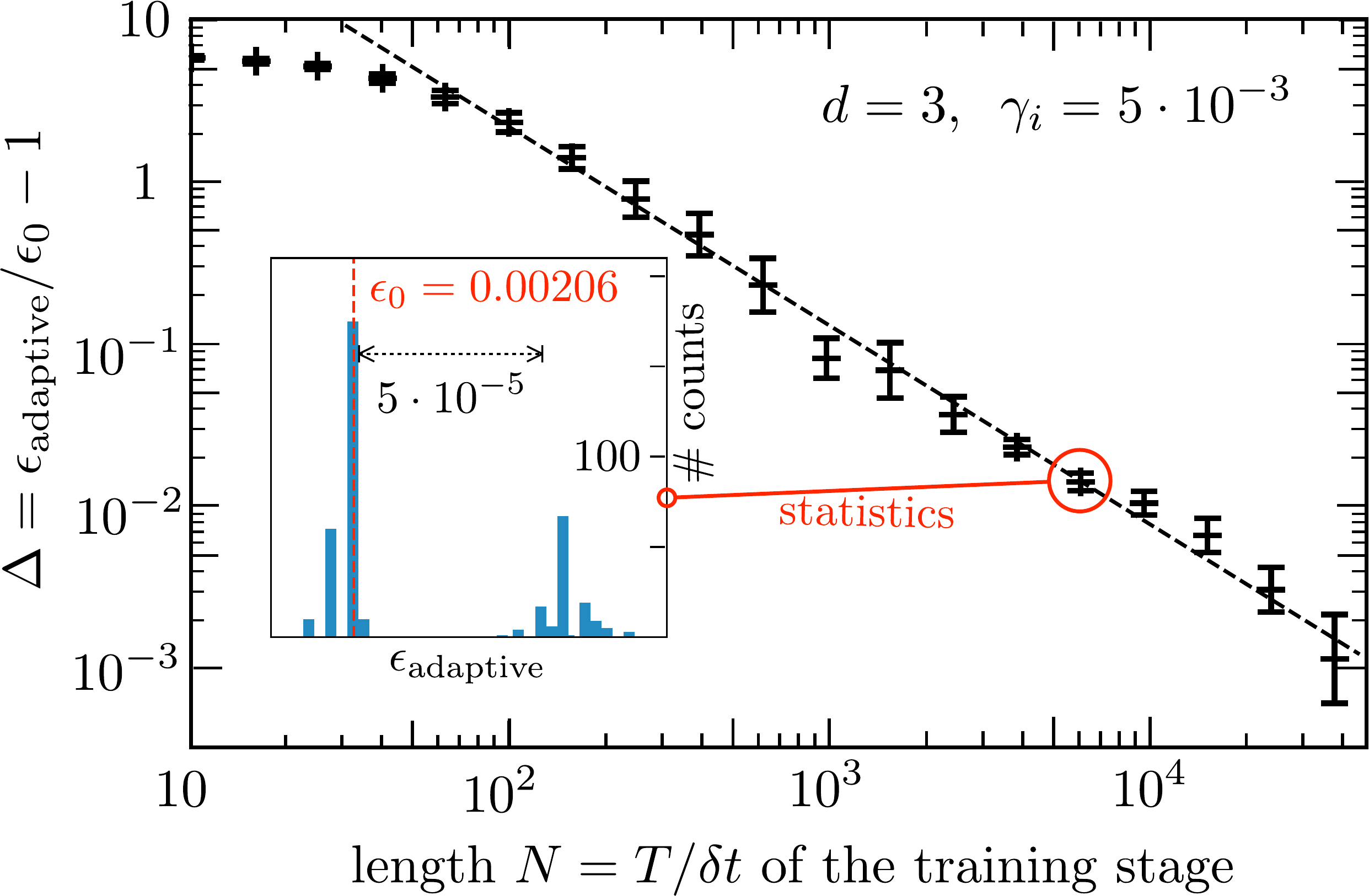}}
\caption{Convergence of the adaptive decoder towards the ideal blossom decoder, as determined by the relative decoder error $\Delta$ as a function of the number of cycles $N$ used to estimate the error probabilities in the training stage. Each data point with error bars results from the repetition of $400$ training stages, the inset shows the statistics for one particular data point. The dashed line through the data points is a guide to the eye.
}
\label{fig:convergence}
\end{figure}

Results are shown in Fig.~\ref{fig:convergence}, for a depth $d=3$ repetition code with uniform single-qubit error rate $\gamma_i=5\cdot 10^{-3}$. We observe a power law convergence $\Delta\propto N^{-\alpha}$ with $\alpha\approx 1.2$. (We do not have an analytical result for this exponent.)

\subsection{Performance in a time-dependent environment}

\begin{figure}[tb]
\centerline{\includegraphics[width=0.9\linewidth]{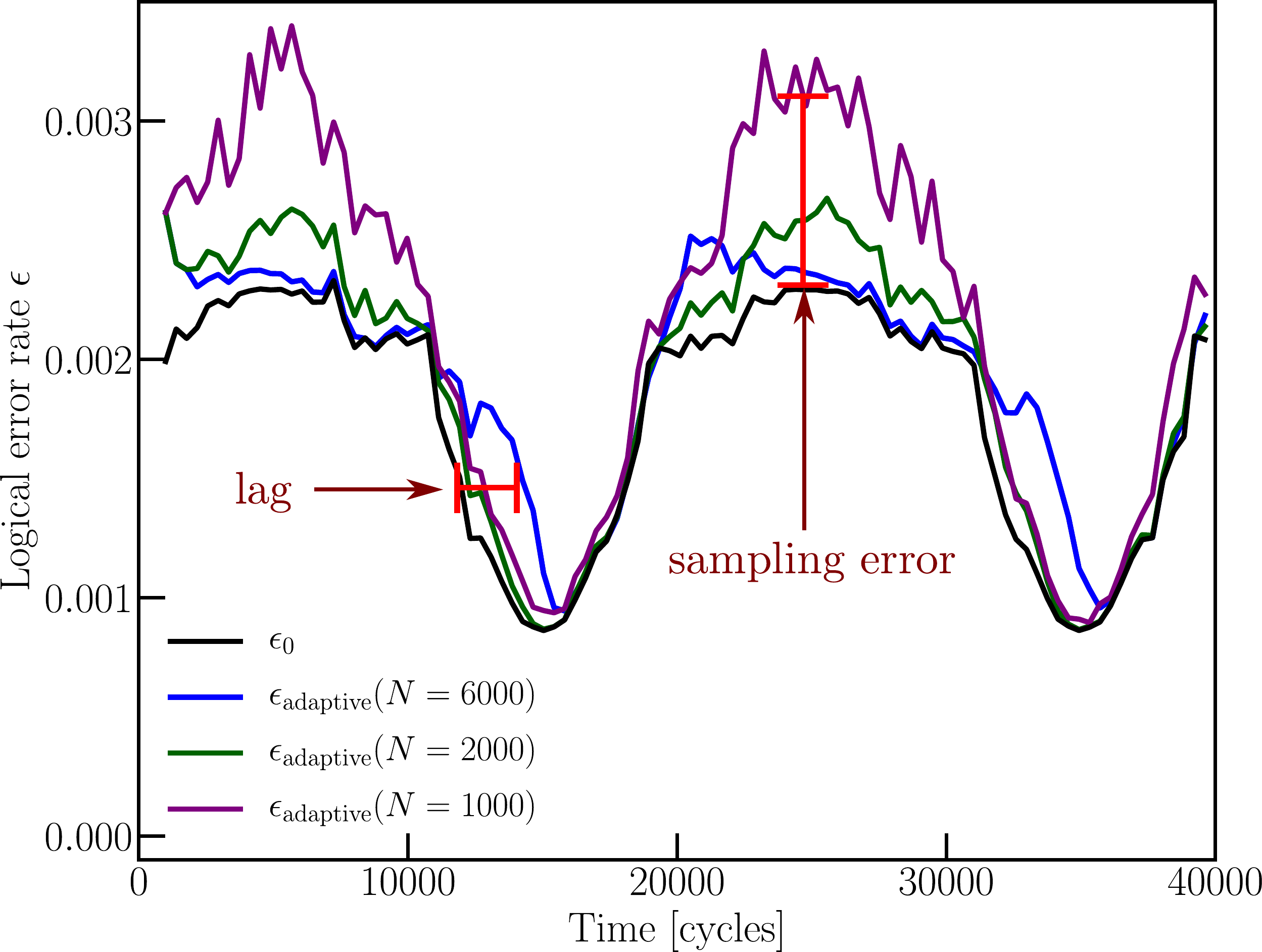}}
\caption{Performance of the adaptive decoder in the presence of a fluctuating noise ($d=3$, $\gamma_i=0.005$ for data qubits, $\gamma_i=0.005+0.005\sin(\pi t/10^4\delta t)$ for ancilla qubits) using three different time windows $T=N\delta t$ for the error estimation. The average over 200 training stages is compared to a blossom decoder (black) with optimally chosen weights at every point in time. Small time windows suffer from sampling error, but adapt quickly to changing error rates, while a decoder with a larger time window lags behind. The optimal time window that balances the two effects is around $T=2000\,\delta t$ in this case.
}
\label{fig:fluctuations}
\end{figure}

The adaptive decoder can be readily applied to sources of noise that vary in time, by recalibration of the weights as time proceeds. We implement this by estimating the error probabilities at time $t$ from the syndrome data in the time interval $(t-T,t)$. The optimal time window $T=N\delta t$ should not be too short in view of the statistical error~\eqref{eq:stderr}, and it should not be too large in view of the variation $\omega T p_{ij}$ of the probabilities in the time-dependent environment (with characteristic frequency $\omega$). The sum of these sources of error is minimized for
\begin{equation}
N_{\rm opt}\simeq(p_{ij}\omega^2\delta t^2)^{-1/3}\Rightarrow\delta p_{ij}^{\rm opt}\simeq p_{ij}^{2/3}(\omega\delta t)^{1/3}\label{eq:opt_timewindow}.
\end{equation}
The adaptive decoder fails if the noise fluctuates too rapidly to acquire sufficient data for the probability estimation. The condition $\delta p_{ij}^{\rm opt}\ll p_{ij}$ implies an upper bound 
\begin{equation}
\omega_c\simeq \bar{p}/\delta t
\end{equation}
on the frequency of the noise variations that is adaptable for a typical error probability $\bar{p}$.

We test the adaptive decoder in the presence of time dependent errors by taking $\gamma_i=\gamma_0$ for the data qubits and $\gamma_i=\gamma_0(1+\sin\omega t)$ for the ancilla qubits (with $\gamma_0=5\cdot 10^{-3}$ and $2\pi/\omega=2\cdot 10^4\,\delta t$). The predicted optimal time window at this frequency, for $\bar{p}=5\cdot10^{-3}$, is $N_{\rm opt}\approx1265$. As shown in Fig.\ \ref{fig:fluctuations}, when a larger window $N\gg N_{\rm opt}$ is used, the decoder experiences a time lag in determining optimal weights; for a smaller window $N<N_{\rm opt}$ the weight estimation is degraded by sampling errors.

\section{Conclusion}
We have demonstrated that it is possible to analytically calculate the underlying error probabilities from measured error syndromes in a broad class of stabilizer codes. As this requires inverting a set of non-linear equations, it is surprising that it should be possible at all, let alone with such small overhead. Because the inversion is exact, the convergence of our adaptive decoder to the ideal blossom decoder should be optimal in the absence of additional information about the error rates. This implies that fluctuations faster than a critical frequency $\omega_c$ are uncorrectable; we have estimated $\omega_c\simeq \bar{p}/\delta t$, with $\bar{p}$ the single-qubit error probability and $\delta t$ the duration of one error-correction cycle. Such rapid fluctuations will contribute relatively more to the logical error rate of a quantum error correcting code than slow fluctuations to which the decoder can adapt.

It would be interesting for future work to test the adaptive decoder on more complex noise models, where the optimal window must be chosen for an entire noise frequency spectrum, instead of for a single frequency. We expect white noise to be significantly worse for quantum error correction than $1/f$ noise, due to the much larger contributions from high frequencies. Future work could also extend our results to simulations of the surface code or topological cluster states.

\acknowledgments

This research was supported by the Netherlands Organization for Scientific Research (NWO/OCW) an ERC Synergy Grant, and by the Office of the Director of National Intelligence (ODNI), Intelligence Advanced Research Projects Activity (IARPA), via the U.S. Army Research Office grant W911NF-16-1-0071. The views and conclusions contained herein are those of the authors and should not be interpreted as necessarily representing the official policies or endorsements, either expressed or implied, of the ODNI, IARPA, or the U.S. Government. The U.S. Government is authorized to reproduce and distribute reprints for Governmental purposes notwithstanding any copyright annotation thereon.

\end{document}